\documentstyle[bezier]{espart}

\newcommand{\beqa}{\begin{eqnarray}}
\newcommand{\eeqa}{\end{eqnarray}}
\newcommand{\beq}{\begin{equation}}
\newcommand{\eeq}{\end{equation}}
\newcommand{\non}{\nonumber}

\begin{document}
\input FEYNMAN

\begin{frontmatter}

\title{BRST treatment of the Bohr collective hamiltonian at high spins}
\author{J. P. Garrahan\thanksref{conicet_jpg}} and
\author{D. R. Bes\thanksref{conicet_drb}}
\address{Departamento de F\'{\i}sica, C.N.E.A., Av. del Libertador
8250, (1429) Buenos Aires, Argentina}
\thanks[conicet_jpg]{Fellow of the CONICET, Buenos Aires, Argentina.
E-mail: garrahan@tandar.edu.ar.}
\thanks[conicet_drb]{Fellow of the CONICET, Buenos Aires, Argentina.
E-mail:  bes@cnea.edu.ar.}

\begin{abstract}
The BRST treatment of triaxial systems rotating at high spins is used
to solve perturbatively the $\gamma$-independent Bohr collective
hamiltonian.
\end{abstract}

\end{frontmatter}

\section{Introduction}

The standard model used to calculate nuclei at high spins is the
cranking model \cite{Go76}.  It is particularly successful in the
prediction of energy levels, single-particle states, deformation
parameters, etc.\ . The problem of constructing the normal modes
associated with the cranking model has been treated by several authors
\cite{Ma77}.  All these treatments assume the quadrupole plus monopole
pairing interaction.  A different treatment of the cranking model at
high spins \cite{KBC90} is based on the application of the BRST symmetry
\cite{BRST,HT92} to many-body problems \cite{BK90}. Since the
collective coordinates are explicitly introduced, the treatment
requires constraints and gauge conditions. An extension of the
formalism to include transition probabilities has recently appeared
\cite{GB94}.

In the BRST treatment there are no restrictions concerning the
hamiltonian\footnote{In particular, the velocity-dependent residual
interactions may be treated on the same footing as the interactions
which are directly responsible for the mean field. The importance of
terms of this type insuring the Galilean invariance of the
hamiltonian has been recently stressed in refs.\ \cite{FL}.}, other
than those associated with the validity of the cranking model:  small
fluctuations in physical magnitudes relative to their expectation value
(if this one does not vanish).  In particular, this must be true for
the collective angular momentum components $\vec{I}$ relative to the
expectation value $\langle I_x \rangle=I$.

None of the previous formalisms has been carried out beyond the level
of constructing normal modes and their corresponding transition matrix
elements to the ground state. However (perturbative) calculations
should also be performed within such basis. In the present letter we
report for the first time on one of such calculations. It is performed
by applying the formalism to a reasonably simple but non-trivial
system, the Bohr collective hamiltonian \cite{Bo52}.  Our aim is: i) to
verify that the problem of infrared divergencies has been properly
solved and ii) to compare the perturbative results in the cranked basis
with those in the laboratory system (which can be solved in this case).
Emphasis is put in those diagrams associated with the modification of
the rotational energies in excited states, since their systematic
application may shed light on the still unsolved problem of identical
bands.

The Bohr collective hamiltonian is constructed out of the five
components of a quadrupole tensor $\alpha_{\mu}$ and their associated
conjugate momenta $p_{\mu}$. We prefer to work with the ``cartesian''
components $q_{\nu}$
\beqa
\alpha_o=q_{\beta} \; ; \;\;\;\;\;
        \alpha_{\pm 1} = \pm \sqrt{\half} (q_y \pm \mathrm{i} q_x)
        \; ; \;\;\;\;\;
        \alpha_{\pm 2} = \sqrt{\half} (q_{\gamma} \mp \mathrm{i} q_z)
        \label{b1}
\eeqa

In terms of such coordinates and momenta, the kinetic energy and the
two (independent) static invariants of the problem may be written
\beqa
T &=& \half (p_{\beta}^2+p_{\gamma}^2+p_x^2+p_y^2+p_z^2) \\
W_2 &=&  q_{\beta}^2+q_{\gamma}^2+q_x^2+q_y^2+q_z^2 \equiv \beta^2 \\
W_3 &=& - 2 q_{\beta}^3 + 6 q_{\beta} q_{\gamma}^2 +
        3 q_{\beta} (2 q^2_z - q^2_x - q^2_y) +
        3 \sqrt{3} q_{\beta} (q_x^2 - q_y^2) \nonumber \\
        & & - 6 \sqrt{3} q_x q_y q_z \equiv - 2 \beta^3 \cos{3 \gamma}
	\label{b2}
\eeqa

We work in a rotating frame of reference, and we formulate the
problem within an overcomplete space consisting of the five
original\footnote{ These five original degrees of freedom of the Bohr
hamiltonian are usually supposed to represent collective coordinates.
Note that this is not the interpretation here, since we reserve the
name {\em collective} (as opposed to {\em intrinsic}) to the Euler
angles determining the position of the rotating frame.} or {\em
intrinsic} variables $q_{\nu}$ (and their conjugate momenta $p_{\nu}$)
plus three { \em collective} coordinates, the Euler angles $\phi_{\nu}$
(and the three components of the collective angular momentum in the
moving frame $I_{\nu}$, $[I_{\mu},I_{\nu}]=-\mathrm{i}\epsilon_{\mu \nu
\rho}I_{\rho}$). The components of the intrinsic angular momentum
are given in appendix \ref{AppAngMom}.

Since the collective variables are raised to the level of true
variables, some combination of the intrinsic variables belong to the
spurious sector, which is completed through the introduction of three
Lagrange multipliers $\Omega_{\nu}$ (and their conjugate momenta
$P_{\nu}$) and six ghost (fermion) variables $\eta_{\nu},{\bar
\eta}_{\nu}$ (and their conjugate partners $\pi_{\nu},{\bar
\pi}_{\nu}$).

The BRST hamiltonian for an ellipsoidal system reads
\begin{eqnarray}
H_{\mathrm{BRST}} &=& H_{\mathrm{B}} -
        \Omega_{\nu} (J_{\nu} - I_{\nu}) +
        \frac{1}{F_{\nu}}G_{\nu}P_{\nu} -
        \frac{A_{\nu}}{2 F_{\nu}^2} P_{\nu}^2 \nonumber\\
        & & + \mathrm{i} \pi_{\nu} {\bar \pi}_{\nu} -
        \frac{1}{F_{\nu}}  {\bar \eta}_{\mu}
        \eta_{\nu}  [G_{\mu},J_{\nu}] +
        \mathrm{i}  \epsilon_{\mu \nu \rho}
        \Omega_{\rho} \pi_{\mu} \pi_{\nu} \label{x3}
\end{eqnarray}
where $H_{\mathrm{B}}$ is the original Bohr hamiltonian, and $G_{\mu}$
are gauge-fixing functions which we choose later. Three of the
constants $F_{\nu},A_{\nu}$ disappear during the treatment, while the
remaining three give rise to the spurious frequencies
$\omega_x,\omega_p$ (eqs. (\ref{h2x}) and (\ref{h2yz})).

The coordinates $q_{\beta},q_{\gamma}$ and the $x$-components of the
intrinsic and collective angular momentum and of the Lagrange
multipliers have the non-vanishing expectation values
\beqa
\langle q_{\beta} \rangle &=& \beta_o \cos{\gamma_o}
        \; ; \;\;\;\;\;
        \langle q_{\gamma} \rangle = \beta_o \sin{\gamma_o}
        \label{bz3} \\
\langle J_x \rangle &=& \langle I_x \rangle = \Im \langle \Omega_x
\rangle =I
\label{b3}
\eeqa

The minimization of the hamiltonian $H_{\mathrm{BRST}}$ together with
the condition (\ref{b3}) implies the minimization of the routhian
$\delta \langle (H - \Omega J_x) \rangle =0$ $ (\Omega\equiv \langle
\Omega_x \rangle)$, and yields the value of the deformation parameters
$\beta_o(I),\gamma_o(I)$ and the moment of inertia $\Im(I)$. Therefore
the cranking solution is the starting point of a treatment of
collective rotational motion which is exact in principle.

If the potential energy surface $V=V(W_2,W_3)$ is independent of $W_3$
(i.e., $\gamma$-independent) the minimization condition yields
\beq
\gamma_o=\pi /6 \; ; \;\;\;\;\;\; \Im=4\beta^2_o \; ; \;\;\;\;\;\;
        \langle p_x \rangle = 2 \Omega \beta_o \label{bv5}
\eeq
Since there is no restoring force in the $\gamma$-direction, the system
adquires inmediately the $\gamma_o$ deformation that minimizes the
kinetic energy\footnote{This fact explains the difficulties in deciding
empirically between the $\gamma$-unstable and the spheroidal rotor
descriptions.}.

Because $W_2=\beta^2$, the value of $\beta_o$ is determined from the
eq.\ :
\beq
V_1-\frac{I^2}{8\beta^4_o}=0 \label{mincond}
\eeq
where $V_i \equiv
\frac{d^iV(\beta^2)}{(d\beta^2)^i}|_{\beta^2=\beta_o^2}$. The
expectation value of the BRST hamiltonian is
\beq
\langle H_{\mathrm{BRST}} \rangle =
        \frac{I^2}{2\Im} + V(\beta_o^2) \label{b6}
\eeq

In the next step we construct the normal modes. Each operator may
consist of several terms which are labelled by the number of phonons
through the supraindex ($\iota$) (for instance, $\langle a \rangle =
a^{(0)}$).  The normal modes are determined from the quadratic terms in
the hamiltonian. Due to signature-invariance, these terms split
into two parts which are labelled by the subindices $x$ and $\perp$,
respectively. The expression for $H_x^{(2)}$ is
\beqa
H_{ax}^{(2)} &=& H^{(2)}_{\mathrm{B}x} - \Omega J^{(2)}_{xx}
        \nonumber \\
        &=& \half (p_{\beta}^{(1)2} + p_{\gamma}^{(1)2} + p_x^{(1)2}) +
        V_1 (q^{(1)2}_{\beta} + q^{(1)2}_{\gamma} + q^{(1)2}_x)
        \nonumber \\
        & & + \half V_2 \beta_o^2
        (\sqrt{3} q_{\beta}^{(1)} + q_{\gamma}^{(1)})^2-
        \Omega J^{(2)}_{xx} \label{wqa}\\
H^{(2)}_{bx} &=& - \Omega_x^{(1)} J_x^{(1)} +
        \frac{1}{F_x} G_x P_x - \frac{A_x}{2F_x^2} P_x^2 +
        \mathrm{i} \pi_x \bar{\pi}_x +
        \frac{\mathrm{i}}{F_x} \eta_x \bar{\eta}_x  \label{hh22x}\\
H^{(2)}_x &=& H_{ax}^{(2)} + H_{bx}^{(2)}
        = \omega_{\beta}
        (\Gamma_{\beta}^{\dag} \Gamma_{\beta} + \half) +
        \omega_{\gamma}
        (\Gamma_{\gamma}^{\dag} \Gamma_{\gamma} + \half)
        \nonumber \\
        & & + \omega_x (\Gamma_{1x}^{\dag} \Gamma_{1x} -
        \Gamma_{0x}^{\dag} \Gamma_{0x} + {\bar a}_x a_x +
	{\bar b}_x b_x)
        \label{h2x}
\eeqa
The hamiltonian includes the two real vibrations with frequencies
$\omega_{\gamma}=2\Omega$ and $\omega_{\beta} = \sqrt{8 V_1 + \Im V_2}$
and a (supersymmetric) spurious sector arising from (\ref{hh22x})
and from a term $(J^{(1)}_x)^2/2\Im_x$ obtained in (\ref{wqa}). The
existence of this term is a consequence of the fact that the first two
lines commute with the operator $J^{(1)}_x$. The construction of the
spurious sector\footnote{Note the relations
${[\Gamma_{1s},\Gamma^{\dag}_{1s}]}=
-{[\Gamma_{0s},\Gamma^{\dag}_{0s}]}={[{\bar a}_s,a_s]}_+= {[{\bar
b}_s,b_s]}_+=1$ for $s=x,p$.} is made as in \cite{KBC90,BK90,GB94}. The
(arbitrary) frequency $\omega_x$ should disappear from any physical
result. The (linear) gauge $G_x$ is the RPA angle which is obtained
simultaneously with the dynamical moment of inertia from the RPA eqs.\
\cite{MW70}
\beq
{[H_{ax}^{(2)},G_x]} =
        - \frac{\mathrm{i}}{\Im_x} J^{(1)}_x
        \; ; \;\;\;\;\; [G_x,J_x^{(1)}] = \mathrm{i} \label{b12}
\eeq
which yield
\beq
G_x = \frac{1}{4 \beta_o} \left(1 + \frac{\Im}{\Im_x} \right) q_x^{(1)} -
        \frac{4 I \beta_o}{V_2 \Im_x \Im^2}
        (\sqrt{3} p_{\beta}^{(1)} + p_{\gamma}^{(1)})
        \; ; \;\;\;\;\;\;
        \Im_x = \Im + \frac{16 \Omega^2}{V_2}
        \label{bg12}
\eeq

Note the presence of the quadratic term $-\Omega J^{(2)}_{xx}$
(eq.\ (\ref{j2xx})) in the hamiltonian with positive signature
(\ref{wqa}), which has been overlooked in previous derivations of the
normal modes in cranked systems.

The quadratic hamiltonian in the perpendicular direction is
\beqa
H^{(2)}_{a\perp} &=& \half (p_y^{(1)2}+p_z^{(1)2}) +
        V_1 (q^{(1)2}_y + q^{(1)2}_z) -
        \Omega (J^{(2)}_{x\perp}-I^{(2)}_x) \\
H^{(2)}_{b\perp} &=& - \Omega_y (J^{(1)}_y-I_y^{(1)}) +
        \frac{1}{F_y} G_y P_y - \frac{A_y}{2F_y} P^2_y -
        \Omega_z (J^{(1)}_z-I^{(1)}_z) + \frac{1}{F_z} G_z P_z
	\non \\
        & &
        - \frac{A_z}{2F_z} P^2_z \;
        + \mathrm{i} \pi_y \bar{\pi}_y +
        \frac{\mathrm{i}}{F_y} \eta_y \bar{\eta}_y +
        \mathrm{i} \pi_z \bar{\pi}_z +
        \frac{\mathrm{i}}{F_z}\eta_z \bar{\eta}_z +
        \mathrm{i} \Omega (\pi_y\eta_z-\pi_z\eta_y) \\
H^{(2)}_{\perp}&=& H^{(2)}_{a\perp}\,+\,H^{(2)}_{b \perp} \nonumber \\
        &=& \half \Omega +
        \omega_{\mathrm w}
        (\Gamma^{\dag}_{\mathrm{w}} \Gamma_{\mathrm{w}} + \half) +
        \omega_p
        (\Gamma^{\dag}_{1p}\Gamma_{1p} -
         \Gamma^{\dag}_{0p}\Gamma_{0p} +
         \bar{a}_p a_p + \bar{b}_p b_p) \label{h2yz}
\eeqa
where $p=\pm$.  The linear and quadratic terms of the collective
angular momentum are obtained by means of the Holstein-Primakoff
representation \cite{HP40}.  The first (constant) term in (\ref{h2yz})
plus the rotational term in eq.  (\ref{b6}) yield the quantal form
$I(I+1)$. The real degree of freedom in (\ref{h2yz}) represents the
wobbling motion with frequency $\omega_{\mathrm w}=3\Omega$.  There
also appear two spurious sectors with (arbitrary) frequencies
$\omega_p$.  The gauge functions and the moments of inertia are derived
from a generalization of the RPA eqs.\ (\ref{b12}) \cite{KBC90}
\beqa
{[H_{a\perp},G_y]} &=& - \frac{\mathrm i}{\Im_y} J^{(1)}_y +
        \mathrm{i} I \left( \frac{1}{\Im_y}-\frac{1}{\Im} \right) G_z
        \; ; \;\;\;\;\;
        [G_y,J^{(1)}_y]= {\mathrm i} \\
{[H_{a\perp},G_z]}&=& - \frac{i}{\Im_z} J^{(1)}_z -
        \mathrm{i} I \left( \frac{1}{\Im_z}-\frac{1}{\Im} \right) G_y
        \; ; \;\;\;\;\;
        [G_z,J^{(1)}_z]= {\mathrm i}
\eeqa
plus the condition ${[G_y,G_z]}= 0$. These eqs. yield $G_y=-
\beta_o^{-1} q^{(1)}_y$, $G_z=- \beta^{-1}_o q^{(1)}_z$ and $ \Im_y =
\Im_z = \beta^2_o$.

The quadratic hamiltonians (\ref{h2x}) and (\ref{h2yz}) determine both
the real and the spurious spectra associated with a given value of $I$.
The corresponding states span a space which is factorized into real,
spurious and rotational subspaces
\begin{eqnarray}
|n_{\beta},n_{\gamma},n_{\mathrm w} \rangle_I \,
                |n_{0 \nu},n_{1 \nu},n_{a \nu},n_{b \nu}\rangle_I \,
                |IM\rangle =
                {\cal N} \,
                (\Gamma^{\dag}_{\beta})^{n_{\beta}}
                (\Gamma^{\dag}_{\gamma})^{n_{\gamma}}
                (\Gamma^{\dag}_{w})^{n_{\mathrm w}} \,
                |r\rangle_I
                \nonumber \\
                \times
                (\Gamma^{\dag}_{0 \nu})^{n_{0 \nu}} \,
                (\Gamma^{\dag}_{1 \nu})^{n_{1 \nu}} \,
                ({\bar a_{\nu}})^{n_{a \nu}}
                ({\bar b_{\nu}})^{n_{b \nu}} \,
                |sp\rangle_I \,
                |IM\rangle \label{z7}
\end{eqnarray}
where  $n_{\beta},n_{\gamma},n_{\mathrm w},n_{1 \nu},n_{0
\nu}=0,1,2,\ldots$, and $n_{a \nu},n_{b \nu}=0,1$.  The product of the
vacua of the real and spurious excitations ($|\mathrm{r}\rangle_I$ and
$|\mathrm{sp}\rangle_I$, respectively) is also the yrast state for a
given $I$. The rotational states $| IM \rangle$ are, in configuration
space, the wave functions $D_{MI}^I$ of the rigid top.  The rotational
excitations $D^I_{MK}$ ($K<I$) are partly associated with the wobbling
motion and partly with the spurious sector. Here $K$ is the projection
along the intrinsic $x$-axis.

We calculate the matrix elements of the quadrupole operator in the
laboratory frame, since they are physical operators. In order to apply
the formalism developed in \cite{GB94} it is convenient to express the
intrinsic operators with respect to the $x$-axis
\beq
\alpha^{(\mathrm{lab})}_{\mu} = D^2_{\mu \nu}(\phi_{\tau})
        D^2_{\nu \rho} (-\half \pi,-\half \pi,0)
        \alpha_{\rho} \label{b18}
\eeq

If acting within states having $I_x \approx I \gg 1$, the rotational
matrices may be expressed in terms of the collective generators
$I_{\nu}$ times the operators $E^{2\iota}_+$ increasing the angular
momentum of the ground state from $I$ to $I+\iota$ \cite{Ma75}. In such
a way we can label the components of (\ref{b18}) both by the number of
phonons and by the change in $I$. In particular, the zero-phonon
component is
\beq
(\alpha_2^{(\mathrm{lab})})^{(0)}=-E^4_+ \, \beta_o/\sqrt{2}
\label{b19}
\eeq
corresponding to the transition matrix element between  yrast states
(which differ in two units of angular momentum).

The linear transition operators can be constructed from the expressions
for the intrinsic variables in terms of the normal modes
\begin{eqnarray}
({\alpha}^{(\mathrm{lab})}_o)^{(1)} &=& \frac{\beta_o}{\sqrt{I}}
        (\Gamma_{\gamma}^{\dag} + \Gamma_{\gamma}) \;;\;\;\;\;\;\;
({\alpha}^{(\mathrm{lab})}_1)^{(1)} =
        \frac{\mathrm{i} \beta_o}{\sqrt{I}} E_+^2
         \Gamma_{\mathrm w} \nonumber\\
({\alpha}^{(\mathrm{lab})}_2)^{(1)} &=& -
        \frac{1}{2 \omega_{\beta}}
        \sqrt{\frac{1}{\omega_{\beta}}}
        E_+^{4} \left(
        \Gamma_{\beta}^{\dag} (\omega_{\beta} - 4 \Omega) +
        \Gamma_{\beta} (\omega_{\beta} + 4 \Omega)
        \right) \label{b20}
\end{eqnarray}

The classification of levels in terms of the quantum numbers
($n_{\beta}, n_{\mathrm w},n_{\gamma}$) allows for a new interpretation
of the degeneracies associated with the $\gamma$-unstable spectrum: for
each $I$, there is vibrational spectrum with $n_{\gamma}=0,1,2,\ldots$.
After $n_{\gamma}=2$, they become degenerate with levels with one
wobbling phonon more than the corresponding states with $I\pm 1$ (fig
\ref{spectra}).

The vanishing of the zero-phonon terms proportional to $E^{2\iota}_+$
with $\iota=0$ (diagonal matrix elements) and of the matrix elements
increasing in one unit both the angular momentum and the number of
wobbling bosons is due to the consevation of the $\gamma$-parity
quantum number (see \cite{Bes58}).

Perturbative corrections can be calculated straightfordwardly since
$H_{\mathrm{BRST}}$ is free of zero frequency modes. The perturbative
expansion yields the residual hamiltonian $H_{\mathrm{res}} =
H_{\mathrm{BRST}}^{(3)}+H_{\mathrm{BRST}}^{(4)}+\cdots$, with
\beqa
H_{\mathrm{BRST}}^{(3)} &=&
	\beta_o V_2 (\sqrt{3} q_{\beta}^{(1)} + q_{\gamma}^{(1)})
	(q_{\beta}^{(1)2} + q_{\gamma}^{(1)2} + q_x^{(1)2} +
	q_y^{(1)2} + q_z^{(1)2})
	\nonumber \\
	& & +
	\frac{\beta_o^3}{6}
	V_3 (\sqrt{3} q_{\beta}^{(1)} + q_{\gamma}^{(1)})^3 -
        \Omega_x (J_x^{(2)} - I_x^{(2)}) -
	\Omega_y J_y^{(2)} - \Omega_z J_z^{(2)}
	\nonumber \\
	& & -
        \frac{1}{F_{\nu}} {\bar \eta}_{\mu}
	\eta_{\nu}  [G_{\mu},J_{\nu}^{(2)}] +
	\mathrm{i} \epsilon_{\mu \nu \rho}
        \Omega_{\rho} \pi_{\mu} \pi_{\nu} \\
H_{\mathrm{BRST}}^{(4)} &=& \half V_2
	(q_{\beta}^{(1)2} + q_{\gamma}^{(1)2} + q_x^{(1)2} +
	q_y^{(1)2} + q_z^{(1)2})^2
	\nonumber \\
	& & +
	\beta_o^2 V_3 (\sqrt{3} q_{\beta}^{(1)} + q_{\gamma}^{(1)})^2
	(q_{\beta}^{(1)2} + q_{\gamma}^{(1)2} + q_x^{(1)2} +
	q_y^{(1)2} + q_z^{(1)2})
	\nonumber \\
	& & +
	\frac{\beta_o^4}{24}
	V_4 (\sqrt{3} q_{\beta}^{(1)} + q_{\gamma}^{(1)})^4 +
	\Omega_y I_y^{(3)} + \Omega_z I_z^{(3)}
\eeqa
The $J_{\nu}^{(2)}$ are given in appendix \ref{AppAngMom} and
$I_{\nu}^{(i)}$ are obtained by means of the Holstein-Primakoff
representation.

\begin{figure}
\begin{picture}(36000,8000)(8000,0)
\drawline\fermion[\NW \REG](12000,2000)[4000]
\global\advance \pmidx by -1200
\put(\pmidx,\pmidy){$x$}
\drawline\fermion[\NE \REG](\pfrontx,\pfronty)[4000]
\global\advance \pmidx by 500
\put(\pmidx,\pmidy){$x$}
\drawline\fermion[\N  \REG](\pfrontx,\pfronty)[4000]
\put(\pmidx,\pmidy){$x$}
\put(\pfrontx,\pfronty){\circle*{300}}
\drawline\fermion[\NW \REG](20000,2000)[4000]
\global\advance \pmidx by -1700
\put(\pmidx,\pmidy){$\perp$}
\drawline\fermion[\NE \REG](\pfrontx,\pfronty)[4000]
\global\advance \pmidx by 800
\put(\pmidx,\pmidy){$\perp$}
\drawline\fermion[\N  \REG](\pfrontx,\pfronty)[4000]
\put(\pmidx,\pmidy){$x$}
\put(\pfrontx,\pfronty){\circle*{300}}
\drawline\fermion[\NW \REG](28000,2000)[4000]
\global\advance \pmidx by -1200
\put(\pmidx,\pmidy){$g$}
\drawline\fermion[\NE \REG](\pfrontx,\pfronty)[4000]
\global\advance \pmidx by 500
\put(\pmidx,\pmidy){$g$}
\drawline\fermion[\N  \REG](\pfrontx,\pfronty)[4000]
\put(\pmidx,\pmidy){$x$}
\put(\pfrontx,\pfronty){\circle*{300}}
\drawline\fermion[\N \REG](34000,2000)[4000]
\put(\pfrontx,\pfronty){\circle*{300}}
\put(\pmidx,\pmidy){$x$}
\drawline\fermion[\S \REG](38000,4000)[3000]
\global\advance \pmidx by 200
\put(\pmidx,\pmidy){$x$}
\drawline\fermion[\N \REG](\pfrontx,\pfronty)[3000]
\global\advance \pmidx by 200
\put(\pmidx,\pmidy){$x$}
\put(37350,3600){$\Large{\times}$}
\put(41350,3600){$\Large{\times}$}
\end{picture}
\begin{picture}(36000,12000)
\put(3350,5600){$\Large{\times}$}
\bezier{200}(12000,2000)(8000,6000)(12000,10000)
\bezier{200}(12000,2000)(16000,6000)(12000,10000)
\drawline\fermion[\N \REG](12000,2000)[8000]
\put(\pmidx,\pmidy){$x$}
\global\advance \pmidx by 2100
\put(\pmidx,\pmidy){$x$}
\global\advance \pmidx by -4800
\put(\pmidx,\pmidy){$x$}
\put(\pfrontx,\pfronty){\circle*{300}}
\put(\pbackx,\pbacky){\circle*{300}}
\bezier{200}(20000,2000)(16000,6000)(20000,10000)
\bezier{200}(20000,2000)(24000,6000)(20000,10000)
\drawline\fermion[\N \REG](20000,2000)[8000]
\put(\pmidx,\pmidy){$x$}
\global\advance \pmidx by 2100
\put(\pmidx,\pmidy){$\perp$}
\global\advance \pmidx by -5000
\put(\pmidx,\pmidy){$\perp$}
\put(\pfrontx,\pfronty){\circle*{300}}
\put(\pbackx,\pbacky){\circle*{300}}
\bezier{200}(28000,2000)(24000,6000)(28000,10000)
\bezier{200}(28000,2000)(32000,6000)(28000,10000)
\drawline\fermion[\N \REG](28000,2000)[8000]
\put(\pmidx,\pmidy){$x$}
\global\advance \pmidx by 2100
\put(\pmidx,\pmidy){$g$}
\global\advance \pmidx by -4800
\put(\pmidx,\pmidy){$g$}
\put(\pfrontx,\pfronty){\circle*{300}}
\put(\pbackx,\pbacky){\circle*{300}}
\drawline\fermion[\N \REG](34000,2000)[8000]
\put(\pmidx,\pmidy){$x$}
\put(\pfrontx,\pfronty){\circle*{300}}
\put(\pbackx,\pbacky){\circle*{300}}
\end{picture}
\begin{picture}(36000,12000)
\bezier{200}(2000,10000)(3000,4000)(6000,2000)
\bezier{200}(6000,2000)(5000,8000)(2000,10000)
\drawline\fermion[\S \REG](2000,10000)[8000]
\global\advance \pmidx by -1200
\put(\pmidx,\pmidy){$\beta$}
\global\advance \pmidx by 2500
\global\advance \pbacky by 1000
\put(\pmidx,\pbacky){$x$}
\put(\pfrontx,\pfronty){\circle*{300}}
\drawline\fermion[\N \REG](6000,2000)[8000]
\global\advance \pmidx by 200
\put(\pmidx,\pmidy){$\beta$}
\global\advance \pmidx by -2200
\global\advance \pbacky by -1500
\put(\pmidx,\pbacky){$x$}
\put(\pfrontx,\pfronty){\circle*{300}}
\bezier{200}(10000,10000)(11000,4000)(14000,2000)
\bezier{200}(14000,2000)(13000,8000)(10000,10000)
\drawline\fermion[\S \REG](10000,10000)[8000]
\global\advance \pmidx by -1200
\put(\pmidx,\pmidy){$\beta$}
\global\advance \pmidx by 2500
\global\advance \pbacky by 1000
\put(\pmidx,\pbacky){$\perp$}
\put(\pfrontx,\pfronty){\circle*{300}}
\drawline\fermion[\N \REG](14000,2000)[8000]
\global\advance \pmidx by 200
\put(\pmidx,\pmidy){$\beta$}
\global\advance \pmidx by -2200
\global\advance \pbacky by -1500
\put(\pmidx,\pbacky){$\perp$}
\put(\pfrontx,\pfronty){\circle*{300}}
\bezier{200}(18000,10000)(19000,4000)(22000,2000)
\bezier{200}(22000,2000)(21000,8000)(18000,10000)
\drawline\fermion[\S \REG](18000,10000)[8000]
\global\advance \pmidx by -1200
\put(\pmidx,\pmidy){$\beta$}
\global\advance \pmidx by 2500
\global\advance \pbacky by 1000
\put(\pmidx,\pbacky){$g$}
\put(\pfrontx,\pfronty){\circle*{300}}
\drawline\fermion[\N \REG](22000,2000)[8000]
\global\advance \pmidx by 200
\put(\pmidx,\pmidy){$\beta$}
\global\advance \pmidx by -2200
\global\advance \pbacky by -1500
\put(\pmidx,\pbacky){$g$}
\put(\pfrontx,\pfronty){\circle*{300}}
\drawline\fermion[\SW \REG](28000,6000)[4000]
\global\advance \pmidx by -1200
\put(\pmidx,\pmidy){$x$}
\put(\pbackx,\pbacky){\circle*{300}}
\drawline\fermion[\S \REG](\pfrontx,\pfronty)[4000]
\global\advance \pmidx by 200
\put(\pmidx,\pmidy){$\beta$}
\drawline\fermion[\N \REG](\pfrontx,\pfronty)[4000]
\put(\pfrontx,\pfronty){\circle*{300}}
\global\advance \pmidx by 200
\put(\pmidx,\pmidy){$\beta$}
\drawline\fermion[\S \REG](34000,6000)[4000]
\global\advance \pmidx by 200
\put(\pmidx,\pmidy){$\beta$}
\drawline\fermion[\N \REG](\pfrontx,\pfronty)[4000]
\global\advance \pmidx by 200
\put(\pmidx,\pmidy){$\beta$}
\put(33350,5600){$\Large{\times}$}
\end{picture}
\caption{The vertices and diagrammatic contributions to the energies of
the yrast and one $\beta$-phonon state. Here $x$($\perp$) stands for
$\beta,\gamma,1_x,1_x$($\mathrm{w},1_p,0_p$) phonons and $g$ for
ghosts. The isolated crosses denote the expectation value of
$H_{\mathrm{BRST}}^{(4)}$.}

\label{diags}
\end{figure}
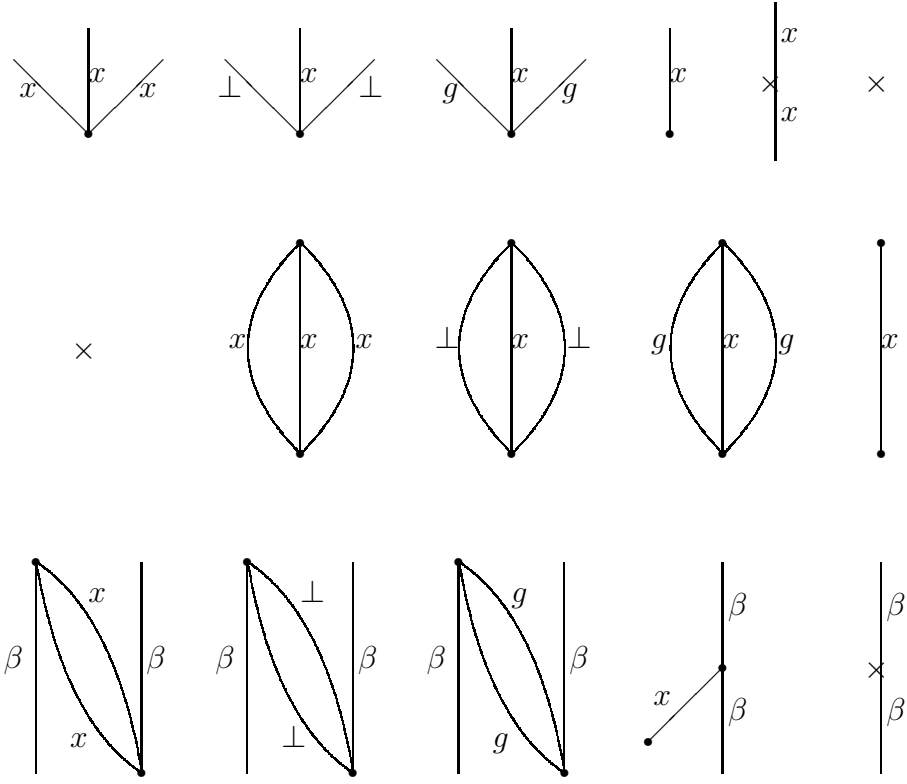

$H_{\mathrm{BRST}}^{(3)}$ and $H_{\mathrm{BRST}}^{(4)}$ give rise to
the vertices shown if fig.\ \ref{diags}. There are no vertices with an
odd number of $\perp$-phonons due signature-invariance, and there are
no vertices with an odd number of ghosts due to the fact that
$H_{\mathrm{BRST}}$ is an even-Grassman function.

The diagrammatic corrections of o($\beta_o^2$) to the energies of the
yrast state and of the one $\beta$-phonon state are given in the second
and third rows of fig.\ \ref{diags}, respectively. Although individual
diagrams depend on the spurious frequencies $\omega_x$ and
$\omega_{\pm}$ this dependence cancels out from the final sums. The
total corrections to the energies are
\begin{eqnarray}
\Delta E(\mbox{yrast state}) &=& \frac{4}{\Im} +
        \frac{18 I}{\omega_{\beta} \Im^2} +
        \frac{3}{\omega_{\beta}^2} \left( - \frac{7}{\Im} V_1 +
        \frac{1}{8} V_2 + \frac{\Im}{8} V_3 +
        \frac{\Im^2}{96} V_4 \right) \nonumber \\
        & & +
        \frac{18 I}{\omega_{\beta}^3 \Im^2} \left( - 8 V_1 +
        \Im V_2 + \frac{\Im^2}{6} V_3 \right) \nonumber \\
        & & -
        \frac{11}{8 \omega_{\beta}^4 \Im} \left( - 8 V_1 +
        \Im V_2 + \frac{\Im^2}{6} V_3 \right)^2 \label{DEyrast} \\
\Delta E(\mbox{one $\beta$-phonon}) &=&
        \frac{36 I}{\omega_{\beta} \Im^2} +
        \frac{3}{\omega_{\beta}^2} \left(\frac{20}{\Im} V_1 +
        \frac{1}{2} V_2 + \frac{\Im}{2} V_3 +
        \frac{\Im^2}{24} V_4 \right) \nonumber \\
        & & +
        \frac{36 I}{\omega_{\beta}^3 \Im^2} \left( - 8 V_1 +
        \Im V_2 + \frac{\Im^2}{6} V_3 \right) \nonumber \\
        & & -
        \frac{15}{2 \omega_{\beta}^4 \Im} \left( - 8 V_1 +
        \Im V_2 + \frac{\Im^2}{6} V_3 \right)^2 \label{DEone}
\end{eqnarray}

The $\gamma$-independent Bohr collective hamiltonian may be
analitically solved in the laboratory frame through the conventional
method of separation of variables \cite{Bes58}. Within such a framework
(which does not require the treatment of a spurious sector) corrections
(\ref{DEyrast}) and (\ref{DEone}) can be reproduced analitically.

In the case of the harmonic motion \cite{Bo52}, $V_1=\mathrm{constant}$
and $V_2=0$. Consequently, the ``static'' moment of inertia is
$\Im=2I/\sqrt{2 V_1}$, while the ``dynamical'' is $\Im_x \rightarrow
\infty$, as required to reproduce yrast energies increasing linearly
with $I$.

A similiar treatment can be applied to a $\gamma$-dependent potential,
the main difference being that $\gamma_o=\gamma_o(I)$ varies from the
equilibrium value at $I=0$ to the value (\ref{bv5}) for large $I$. In
this case the solution requires a numerical computation.

We have perfomed (to our knowledge, for the first time) a perturbative
correction to zero order results appearing through the application of
the cranking approximation. This has been done for the Bohr collective
hamiltonian with a $\gamma$-independent potential energy surface.
Although the empirical applicability of such a model is being
increasingly recognized, we mostly consider the present contribution as
a non-trivial exercise for future calculations with realistic systems.
In particular, the replacement of the initial and final phonons by a
fermion line and the corresponding modification of diagrams measures
the difference between the moments of inertia associated with the even
and odd systems (the problem of identical bands).

\begin{figure}
\begin{picture}(37000,29000)(-2000,-2000)
\put(9500,6000){\vector(4,-3){3700}}
\put(5000,15000){\vector(3,-4){9000}}
\put(28000,21000){\vector(-1,-2){8800}}
\put(17000,6000){\vector(0,-1){3000}}
\thinlines
\multiput(0,0)(0,3000){6}{\line(1,0){5000}}
\multiput(7000,6000)(0,3000){3}{\line(1,0){5000}}
\multiput(14000,3000)(0,3000){4}{\line(1,0){5000}}
\multiput(21000,9000)(0,3000){2}{\line(1,0){5000}}
\multiput(28000,6000)(0,3000){3}{\line(1,0){5000}}
\put(14000,18000){\line(1,0){5000}}
\put(28000,21000){\line(1,0){5000}}
\thicklines
\put(14000,3000){\vector(-3,-1){9000}}
\put(28000,6000){\vector(-3,-1){9000}}
\bezier{50}(18000,3000)(18000,10500)(18000,18000)
\bezier{25}(18000,3000)(21500,6000)(25000,9000)
\put(1000,-2000){$I-2$}
\put(8000,-2000){$I-1$}
\put(14000,-2000){$I$ (even)}
\put(22000,-2000){$I+1$}
\put(29000,-2000){$I+2$}
\put(0,500){$\scriptstyle{(0,0,0)}$}
\put(0,3500){$\scriptstyle{(0,0,1)}$}
\put(0,6500){$\scriptstyle{(0,0,2)}$}
\put(0,9500){$\scriptstyle{(0,0,3) \; (0,2,0)}$}
\put(0,12500){$\scriptstyle{(0,0,4) \; (0,2,1)}$}
\put(0,15500){$\scriptstyle{(1,0,0)}$}
\put(7000,6500){$\scriptstyle{(0,1,0)}$}
\put(7000,9500){$\scriptstyle{(0,1,1)}$}
\put(7000,12500){$\scriptstyle{(0,1,2)}$}
\put(14000,3500){$\scriptstyle{(0,0,0)}$}
\put(14000,6500){$\scriptstyle{(0,0,1)}$}
\put(14000,9500){$\scriptstyle{(0,0,2)}$}
\put(14000,12500){$\scriptstyle{(0,0,3) \; (0,2,0)}$}
\put(14000,18500){$\scriptstyle{(1,0,0)}$}
\put(21000,9500){$\scriptstyle{(0,1,0)}$}
\put(21000,12500){$\scriptstyle{(0,1,1)}$}
\put(28000,6500){$\scriptstyle{(0,0,0)}$}
\put(28000,9500){$\scriptstyle{(0,0,1)}$}
\put(28000,12500){$\scriptstyle{(0,0,2)}$}
\put(28000,21500){$\scriptstyle{(1,0,0)}$}
\end{picture}
\caption{The spectrum of $H_{\mathrm{BRST}}^{(2)}$. The levels are
classified by $(n_{\beta},n_{\mathrm w},n_{\gamma})$. Dark lines
correspond to strong transitions between yrast states, lighter ones to
transitions to vibrational states, and dotted lines to $\gamma$-parity
inhibited transitions.}
\label{spectra}
\end{figure}

\appendix

\section{Intrinsic angular momentum operators} \label{AppAngMom}
The intrinsic angular momentum operators are defined by $J_{\mu} \equiv
\mathrm{i} \sqrt{10} (q,p)_{\mu}^1 $. In cartesian components they are
\begin{eqnarray}
J_x &=& q_z p_y - q_y p_z + (\sqrt{3} q_{\beta} + q_{\gamma}) p_x -
	q_x (\sqrt{3} p_{\beta} + p_{\gamma}) \\
J_y &=& q_x p_z - q_z p_x - (\sqrt{3} q_{\beta} - q_{\gamma}) p_y +
	q_y (\sqrt{3} p_{\beta} - p_{\gamma}) \\
J_z &=& q_y p_x - q_x p_y - 2 q_{\gamma} p_z + 2 q_z p_{\gamma}
\end{eqnarray}

The expansion of the intrinsic coordinates and momenta gives
\begin{eqnarray}
\langle J_x \rangle &=& I \; ; \;\;\;\;\;
	\langle J_y \rangle = \langle J_z \rangle = 0 \\
J_x^{(1)} &=& 2 \beta_o p_x^{(1)} + \frac{I}{2 \beta_o}
	(\sqrt{3} q_{\beta}^{(1)} + q_{\gamma}^{(1)}) \label{jx1} \\
J_y^{(1)} &=& -\beta_o p_y^{(1)} - \frac{I}{2 \beta_o} q_z^{(1)}
	\; ; \;\;\;\;\;
	J_z^{(1)} = -\beta_o p_z^{(1)} + \frac{I}{2 \beta_o} q_y^{(1)} \\
J_x^{(2)} &=& J_{xx}^{(2)} + J_{x\perp}^{(2)}\\
J_{xx}^{(2)} &=&
	(\sqrt{3} q_{\beta}^{(1)} + q_{\gamma}^{(1)}) p_x^{(1)} -
	q_x (\sqrt{3} p_{\beta}^{(1)} + p_{\gamma}^{(1)})
	\label{j2xx} \\
J_{x\perp}^{(2)} &=& q_z^{(1)} p_y^{(1)} - q_y^{(1)} p_z^{(1)} \\
J_y^{(2)} &=& q_x^{(1)} p_z^{(1)} - q_z^{(1)} p_x^{(1)} -
	(\sqrt{3} q_{\beta}^{(1)} - q_{\gamma}^{(1)}) p_y^{(1)} +
	q_y^{(1)} (\sqrt{3} p_{\beta}^{(1)} - p_{\gamma}^{(1)}) \\
J_z^{(2)} &=& q_y^{(1)} p_x^{(1)} - q_x^{(1)} p_y^{(1)} -
	2 q_{\gamma}^{(1)} p_z^{(1)} + 2 q_z^{(1)} p_{\gamma}^{(1)}
\end{eqnarray}

\end{document}